\documentstyle[epsfig,times]{aa}

\def\ltsima{$\; \buildrel < \over \sim \;$}
\def\simlt{\lower.5ex\hbox{\ltsima}} % < over ~
\def\gtsima{$\; \buildrel > \over \sim \;$}
\def\simgt{\lower.5ex\hbox{\gtsima}} % > over ~

\makeatletter
\def\@cite#1#2{(#1\if@tempswa , #2\fi)}
\def\@citex[#1]#2{\if@filesw\immediate\write\@auxout{\string\citation{#2}}\fi
  \def\@citea{}\@cite{\@for\@citeb:=#2\do
    {\@citea\def\@citea{;\penalty\@m\ }\@ifundefined
       {b@\@citeb}{{\bf ?}\@warning
       {Citation `\@citeb' on page \thepage \space undefined}}%
\hbox{\csname b@\@citeb\endcsname}}}{#1}}
\makeatother

\title{BeppoSAX Detection and Follow-up of GRB980425}

\author{E. Pian\inst{1}
\and L. Amati\inst{1}
\and L. A. Antonelli\inst{2}
\and R. C. Butler\inst{3}
\and E. Costa\inst{4}
\and G. Cusumano\inst{5}
\and J. Danziger\inst{6}
\and M. Feroci\inst{4}
\and F. Fiore\inst{7}
\and F. Frontera\inst{1,8}
\and P. Giommi\inst{7}
\and N. Masetti\inst{1}
\and J.M. Muller\inst{7}
\and T. Oosterbroek\inst{9}
\and A. Owens\inst{9}
\and E. Palazzi\inst{1}
\and L. Piro\inst{4}
\and A. Castro-Tirado\inst{10}
\and A. Coletta\inst{7}
\and D. Dal Fiume\inst{1}
\and S. Del Sordo\inst{5}
\and J. Heise\inst{11}
\and L. Nicastro\inst{5}
\and M. Orlandini\inst{1}
\and A. Parmar\inst{9}
\and P. Soffitta\inst{4}
\and V. Torroni\inst{7}
\and J.J.M. in 't Zand\inst{11}
}

\institute{Istituto Te.S.R.E., CNR, via Gobetti 101, I-40129 Bologna, Italy
\and
Osservatorio Astronomico di Roma, sede di Monteporzio Catone, Via Frascati 33,
00040 Monteporzio Catone, Italy
\and
Agenzia Spaziale Italiana, Viale Regina Margherita 120, Roma, Italy
\and 
I.A.S., C.N.R., Via Fosso del Cavaliere, Area della Ricerca di Tor Vergata, 
I-00131 Rome, Italy
\and
I.F.C.A.I., CNR, via Ugo La Malfa 153, I-90146 Palermo, Italy
\and
Osservatorio Astronomico di Trieste, Via G.B. Tiepolo 11, I-34131 Trieste,
Italy
\and
BeppoSAX Scientific Data Center, Via Corcolle 19, I-00131 Rome, Italy
\and
Dip. di Fisica, Univ. di Ferrara, Via Paradiso 11, I-44100 Ferrara, Italy
\and
Astrophysics Division, SSD of ESA, ESTEC, P.O. Box 299, 2200 AG Noordwijk, The
Netherlands
\and
IAA-CSIC, Granada, Spain and LAEFF-INTA, Madrid, Spain
\and
Space Research Organization Netherlands, Sorbonnelaan 2, 3584 CA
Utrecht, The Netherlands
}

\offprints{pian@tesre.bo.cnr.it}
\date{Received \today; Accepted \dots}
\thesaurus{13.07.1,08.19.5: SN~1998bw}

\begin{document}

% This will write DRAFT across each page
%\special{!userdict begin /bop-hook{gsave 200 30 translate
%65 rotate /Times-Roman findfont 220 scalefont setfont
%0 0 moveto 0.88 setgray (DRAFT) show grestore}def end}

\maketitle

\begin{abstract}

We present BeppoSAX GRBM and WFC light curves of GRB980425 and
NFI follow-up data 
%obtained with NFI Target of Opportunity observations
taken in 1998 April, May, and November.  The first NFI observation has
detected within the $8^{\prime}$ radius error box of the GRB an
X-ray source positionally consistent with the supernova SN~1998bw,
exploded within a day of GRB980425, and a fainter X-ray source, not
consistent with the position of the supernova.  The former source is
detected in the following NFI pointings and exhibits a decline of a
factor of two in six months. If it is associated with SN~1998bw, this
is the first detection of hard X-ray emission from a Type I
supernova. 
%The latter source is no longer detected starting $\sim$40
%hours after the GRB.  Selecting either source as a candidate for the
%GRB counterpart leads to radically different scenarios for the
%origin of the $\gamma$-ray event.
The latter source exhibits only marginally significant variability. 
Based on these data, it is not possible to select either
source as a firm candidate for the GRB counterpart.
%Selecting either source as a candidate for the GRB counterpart leads
%to
%radically different scenarios for the origin of the $\gamma$-ray event
%and the evolution of its afterglow.


\end{abstract}

\section{Introduction}

The GRB of 1998 April 25, detected both by the BeppoSAX GRBM and BATSE
and localized with arcminute accuracy by the BeppoSAX WFC, stands out
for its spatial and temporal coincidence with the optically
and radio exceedingly bright Type Ic supernova SN~1998bw (Galama et
al.  1998;  Kulkarni et al. 1998), in the nearby galaxy ESO 184-G82
($z = 0.0085$).  Since the other GRBs for which a redshift measurement
is available are located at larger distances ($z \simgt 1$) and are
characterized by power-law decaying optical afterglows, in agreement
with the ``classical"  fireball model (Rees \& M\'esz\'aros 1992),
this has raised a debate about a possible association between GRBs and
supernovae.  Following the detection of GRB980425, observations of its
error box with the BeppoSAX NFI have been activated 10 hours, one
week, and six months later.  We present here some results and discuss
their implications in view of the detection of SN~1998bw in the GRB
field.  A detailed presentation will be given in Pian
et al.  (1999).

\section{Data analysis and results}

GRB980425 triggered the BeppoSAX GRBM at 21:49:11 UT, and was
simultaneously detected by the WFC unit 2 (Soffitta et al. 1998). 
The event had a duration of 31 s in the range 40-700 keV and of 40 s
in the range 2-26 keV, and exhibited a single, non structured peak
profile in both bands %, with the indication of a $\sim$4 s soft lag
(Fig.~1). The fluences at $\gamma$- and hard X-ray energies are
($2.8 \pm 0.5$)  $\times 10^{-6}$ erg cm$^{-2}$ and ($1.8 \pm 0.3$) 
$\times 10^{-6}$ erg cm$^{-2}$, respectively. (The Galactic 
absorption in the direction of GRB980425, $N_{HI} = 4 \times
10^{20}$ cm$^{-2}$, is negligible at
energies higher than 2 keV.)  %Strong spectral softening is evident
%during the event (Frontera et al.  1999). No peculiar
%temporal or spectral characteristics are noted in this GRB. %\\
The BeppoSAX NFI were pointed at the $8^{\prime}$ radius error box
determined by the WFC at three epochs starting 10 hours after the
GRB (see Table~1; note that the first pointing has been split in two
parts). The preliminary analysis of the LECS and MECS data of the
first portion of the first pointing shows that inside the WFC error
box, two point-like, previously unknown X-ray sources are detected
with a positional uncertainty of $1^{\prime}.5$: 1SAXJ1935.0-5248
(hereafter S1), at RA = 19h 35m 05.9s and Dec = -52$^{\circ}$
50$^{\prime}$ 03$^{\prime\prime}$, and 1SAXJ1935.3-5252 (hereafter
S2), at RA = 19h 35m 22.9s and Dec =-52$^{\circ}$ 53$^{\prime}$
49$^{\prime\prime}$.  Note that the coordinates distributed by Pian
et al. (1998) have been revised in November 1998 (see to this regard
Piro et al. 1998). The revised position of S1 is consistent within
the uncertainty with the position of the optical and radio supernova
SN~1998bw (Galama et al.  1998; Kulkarni et al. 1998), while the
revised position of S2 is $\sim 4^{\prime}$ away from SN~1998bw, and
therefore inconsistent with it (see Fig. 1 in Galama et al. 1999).%\\
The MECS count rates and upper limits for both sources during the three
pointings are reported in Table~1.  The upper limits have been
estimated by taking into account, besides the normal photon statistics,
also the fact that, at these flux levels, the MECS background may be
dominated by the fluctuations of the cosmic X-ray background.  The
observation of November 1998 (taken about a week after the conclusion
of this Conference) shows a decrease in the X-ray flux of S1 of
approximately a factor of two with respect to the level measured in
April-May and the suggestion of slightly extended X-ray emission around
the source. 
%During the second portion of the first pointing, as well as during
%the successive pointings, S2 is no longer detected according to our
%preliminary analysis.  However, both the detection and variability of S2
%have marginal significance (see Table~1). A refined image analysis is
%underway.
During the second portion of the first pointing, as well as in the November
pointing, S2 is not detected, while it is detected in the May pointing,
at a marginally lower level than in the first observation (see Table~1). 

\begin{center}
\begin{tabular}{rccc}
%\hline
\multicolumn{4}{c}{{\bf Table 1:} Journal of BeppoSAX-MECS
Observations}\\
\hline
\hline
\multicolumn{1}{c}{Date (UT)} & t$^a$ (s) & \multicolumn{2}{c}{Flux$^b$ 
($\times 10^{-3}$ cts s$^{-1}$)}\\
 & & S1 & S2 \\
\hline
%1998 Apr 26.334-27.458 & 37220 & $4.6 \pm 0.6^c$ & $2.4 \pm 0.5$ \\
%     Apr 27.469-28.160 & 21805 & $4.5 \pm 0.7$ & $< 1.8$ \\
%     May 02.605-03.621 & 31975 & $3.0 \pm 0.5$ & $< 1.5$ \\ 
%     Nov 10.754-12.004 & 53122 & $1.8 \pm 0.4$ & $< 1.4$ \\ 
1998 Apr 26.334-27.458 & 37220 & $4.6 \pm 0.6^c$ & $2.4 \pm 0.5$ \\
     Apr 27.469-28.160 & 21805 & $4.5 \pm 0.7$   & $< 2.5$ \\
     May 02.605-03.621 & 31975 & $3.0 \pm 0.5$   & $1.4 \pm 0.5$ \\ 
     Nov 10.754-12.004 & 53122 & $1.8 \pm 0.4$   & $< 2.0$ \\ 
\hline
\multicolumn{4}{l}{$^a$ On source exposure time}\\ 
\multicolumn{4}{l}{$^b$ In the energy range 1.6-10 keV}\\
\multicolumn{4}{l}{$^c$ Uncertainties are at 1-$\sigma$; upper
limits are at 3-$\sigma$}\\
\end{tabular}
\end{center}

\section{Discussion}

The count rates in the first line of Table~1 correspond to 
$F_{2-10 keV} \simeq 3
\times 10^{-13}$ erg s$^{-1}$ cm$^{-2}$ for S1 and to $F_{2-10 keV}
\simeq 1.6 \times 10^{-13}$ erg s$^{-1}$ cm$^{-2}$ for S2. The
following data points show a decay for S1 of a factor of two in
$\sim$6 months.  Assuming, as suggested by the positional
coincidence and by variability, that S1 is associated with SN~1998bw, the
observed variation represents a lower limit on the amplitude of
X-ray variability of SN~1998bw.  In fact, the possible NFI detection
of extended emission indicates that S1 might contain a non
negligible contribution from the host galaxy of the supernova.
%, a face-on spiral galaxy about one tenth of the size of our Galaxy.
This is the first detection of hard X-ray emission from a Type I
supernova. At the distance of SN~1998bw, the luminosity observed in
the range 2-10 keV, $5 \times 10^{40}$ erg s$^{-1}$, is compatible
with the luminosity observed in the 0.1-2.4 keV range for the Type
Ic SN~1994I, the only case of soft X-ray Type I supernova emission so
far detected (Immler et al.  1998).  
%Our observation of variable X-ray emission after a Type I supernova
%explosion is also unprecedented. %\\
If SN~1998bw is the counterpart of GRB980425, the production of
$\gamma$-rays could be accounted for by the explosion of a very massive
star ($\sim 40 M_{\odot}$) and by the subsequent expansion of a
relativistic shock, in which non thermal electrons are radiating
photons of $\sim$100 keV, provided the explosion is asymmetric, i.e. 
the GRB is produced in a relativistic jet (Iwamoto et al. 1998; Woosley
et al. 1998;  see however, Kulkarni et al.  1998).  This raises the
hypothesis that two classes of GRBs might exist, with apparently
indistinguishable high energy characteristics, but with different
progenitors.  %\\
On the other hand, disregarding the extremely low
probability of chance coincidence of GRB980425 and SN~1998bw, one might
consider S2 as the X-ray counterpart candidate of the burst.  
%Assuming a power-law decay between the X-ray flux measured by the WFC
%in the 2-10 keV range during the GRB and the flux measured in the
%first NFI observation, we derive a power-law index of $\sim -1.4$.
%This is consistent with the upper limits determined later on by the
%NFI, and is similar to the index found for X-ray afterglows of previous 
%GRBs. 
Assuming a power-law decay between the X-ray flux measured by the WFC
in the 2-10 keV range during the GRB and the flux measured in the first
NFI observation, we derive a power-law index of $\sim -1.4$.  The X-ray
flux measured in May is however a factor $\sim$10 larger than implied
by the power-law decay.  This behavior is unlike that of previously
observed X-ray afterglows, although it could be still reconciled with
it under the hypothesis of a re-bursting superposed on a ``typical"
power-law decline. 

%Further observations by an X-ray
%instrument with higher sensitivity and spatial resolution than BeppoSAX
%can attain might resolve the open issues about the X-ray counterpart of
%GRB980425. 

\begin{acknowledgements}
We thank the BeppoSAX Mission Planning Team and the BeppoSAX SDC and SOC
personnel for help and support in the accomplishment of this project.

\end{acknowledgements}

\begin{figure}
\vspace{0.5cm}
\epsfig{file=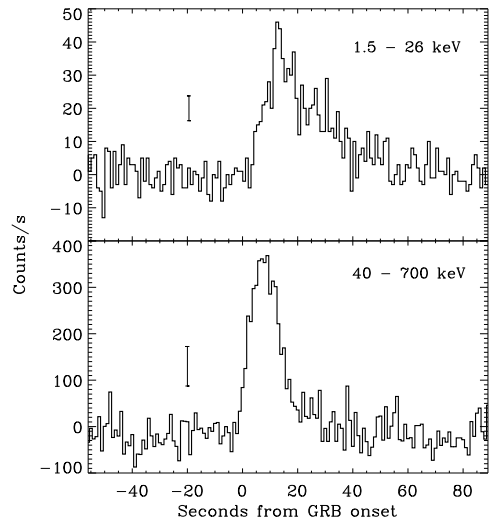,width=7cm}
\label{Fig. 1}
\vspace{0.7cm}
\caption{BeppoSAX WFC (top) and GRBM (bottom) light curves of GRB980425. 
The onset of the GRB, indicated by the zero abscissa, corresponds to 1998
April 25.909097 (i.e., 5 seconds earlier than the GRBM trigger time). The
vertical bars represent the typical error associated with the individual
flux points.}
%\vspace{-1cm}
\end{figure}

\end{document}